\documentclass[english,aps,twocolumn,showpacs,prl,superscriptaddress]{revtex4-1}
\usepackage{pgfplotstable}
\usepackage{pgfplots}
\pgfplotsset{compat=newest}
\pgfplotsset{plot coordinates/math parser=false}
\pgfplotsset{scaled y ticks=false}
\pgfplotsset{every axis title/.append style={at={(0.8,0.1)}}}
\usepackage{bm}

\usepackage{amsmath}
\usepackage{amssymb}

\usepackage{color}

\usepackage{graphicx}
\usepackage{wrapfig}
\usetikzlibrary{pgfplots.groupplots}
\begin{document}

\title{\textit{Ab Initio} Theory of Coherent Laser-Induced Magnetization in Metals}

\author{Marco Berritta}
\email[]{marco.berritta@physics.uu.se}
\affiliation{Department of Physics and Astronomy, Uppsala University, Box 516, SE-75120 Uppsala, Sweden}
\author{Ritwik Mondal}
\affiliation{Department of Physics and Astronomy, Uppsala University, Box 516, SE-75120 Uppsala, Sweden}
\author{Karel Carva}
\affiliation{Department of Physics and Astronomy, Uppsala University, Box 516, SE-75120 Uppsala, Sweden}
\affiliation{Charles University, Faculty of Mathematics and Physics, Department
of Condensed Matter Physics, Ke Karlovu 5, CZ-12116 Prague 2, Czech
Republic}
\author{Peter M.\ Oppeneer}
\affiliation{Department of Physics and Astronomy, Uppsala University, Box 516, SE-75120 Uppsala, Sweden}

\date{\today}

\begin{abstract}
\noindent
{We present the first materials specific \textit{ab initio} theory of the magnetization  
induced by circularly polarized laser light in metals. Our calculations are based on non-linear density matrix 
theory and include the effect of absorption. 
We show that the induced magnetization, commonly referred to as inverse Faraday effect, is strongly materials and frequency dependent, and 
demonstrate the existence of both spin and orbital induced magnetizations which exhibit a surprisingly different behavior. We show that for nonmagnetic metals (as Cu, Au, Pd, Pt) and antiferromagnetic metals the induced magnetization is \textit{antisymmetric} in the light's helicity, whereas for ferromagnetic metals (Fe, Co, Ni, FePt) the imparted magnetization is only asymmetric in the helicity.
We compute effective optomagnetic fields that correspond to the induced magnetizations and provide guidelines for achieving all-optical helicity-dependent switching.}
\end{abstract}

\maketitle

All-optical helicity-dependent magnetization switching  has recently emerged as a promising 
way to manipulate and ultimately control the magnetization in a magnetic material using ultrashort optical laser pulses \cite{kimel05,stanciu07,satoh10,satoh12,kirilyuk13,mangin14,lambert14}.
As femtosecond optical laser pulses are the shortest stimuli known to mankind, all-optical helicity-dependent switching  offers novel options to achieve magnetization reversal at a hitherto unprecedented speed. 
The action of a circularly polarized laser pulse on the magnetization of a material was at first observed for an antiferromagnetic $3d$-metal oxide \cite{kimel05} and later magnetization reversal was demonstrated in a ferrimagnetic rare-earth transition-metal alloy \cite{stanciu07,hassdenteufel13}. Importantly, recent work demonstrated that all-optical helicity-dependent switching is not limited to a special class of materials, but can be achieved in a broader variety of material classes, including metallic multilayers, synthetic ferrimagnets \cite{mangin14}, and even ferromagnets such as FePt \cite{lambert14}, which is the prime candidate material for future ultradense magnetic recording \cite{weller13}.

While these discoveries exemplify that ultrafast magnetization reversal driven by circularly polarized laser pulses could soon revolutionize magnetic recording its underlying physical mechanism is poorly understood. The influence of the circularly polarized laser pulse has been attributed to the inverse Faraday effect (IFE) \cite{kimel05,stanciu07,satoh10,lambert14},  which was discovered fifty years ago \cite{vanderziel65}. The IFE is an optomagnetic counterpart of the magneto-optical Faraday effect, that is, the circularly polarized laser light imparts a magnetization in the material which exerts a torque on the pre-existing magnetization and assists the magnetization switching. However, although various models for the IFE have been proposed \cite{pitaevskii61,pershan66,hertel06,kurkin08,woodford09,popova11,qaiumzadeh13} there does as yet not exist any knowledge as to how the induced magnetization, or optomagnetic field arises, and even less is known about the materials dependence of the IFE. As materials specific theory is lacking it is neither known for which materials large effects are predicted nor how the IFE could be optimized to trigger reversal with minimal laser power.

In this letter we present the first \textit{ab initio} theory of the magnetization optically induced by circularly polarized laser light. Our theoretical framework is based on the materials specific electronic structure as computed within the density-functional theory (DFT) and is applicable to nonabsorbing as well as absorbing materials. Thus, we directly predict the laser-imparted magnetization in ferromagnetic and nonmagnetic metals. We show that the IFE consists of two parts, the spin moment IFE and the orbital moment IFE, which are strongly materials dependent and exhibit a very different behavior. Based on our \textit{ab initio} calculations we provide guidelines for maximal laser-imparted magnetizations.

\begin{figure}
\includegraphics[width=0.8\linewidth]{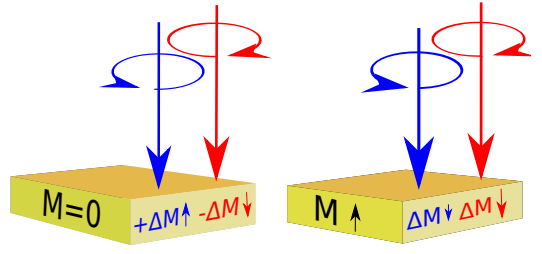}
\caption{(Color online). Schematic view of the helicity-dependent magnetization $\Delta M$ induced by circularly polarized laser light  in materials, (left) with no net magnetization $\bm{M}$ present, and with a net ferromagnetic magnetization (right).
\vspace*{0.3cm}
}
\label{fig:f1}
\end{figure}

We use the recently derived quantum theory for the IFE \cite{battiato14} in which the second-order density response of the electronic states to the external electromagnetic field is evaluated from the Liouville-von Neumann equation. 
The laser-induced magnetization is given by $\bm{M}_{ind} = \mu_B \textrm{Tr} \{ (\hat{\bm{L}} +2 \hat{\bm{S}}) \hat{\rho}^{[2]} \}$, where $\hat{\bm{S}}$ and  $ \hat{\bm{L}}$ are the spin and orbital angular momentum operators and $\hat{\rho}^{[2]}$ is the 2$^{\rm nd}$ order density matrix response. 
For coherent circularly polarized light the magnetization is induced through the photon spin angular momentum along the light's axis, as schematically shown in Fig.\ \ref{fig:f1}. The induced magnetization per unit volume is  \cite{battiato14}
\begin{equation}
\label{eq:formal_ife}
    {M}_{ind}=({\cal K}_{\rm o}+{\cal K}_{\rm dA} +{\cal K}_{\rm dB}+c.c.)E_{0}^2 ,
\end{equation}
where $E_0$ is the electric field amplitude of the incident radiation, chosen to impinge along the $z$-axis, and
\begin{equation}
\begin{split}
{\cal K}_{\rm o} = &
\frac{e^2}{m^2\omega^2} 
\!\! \! \sum_{n\neq m; l}\!\! \! M_{mn}\frac{\frac{p^+_{nl}p^-_{lm}(f_{m}-f_{l})}{E_l-E_m+i\hbar\Gamma_{lm}-\hbar\omega}-\frac{p^-_{nl}p^+_{lm}(f_{l}-f_{n})}{E_n-E_l+i\hbar\Gamma_{nl}-\hbar\omega}}{E_n-E_m+i\hbar\Gamma_{nm}},
\nonumber
\end{split}
\end{equation}
\begin{equation}
\begin{split}
{\cal K}_{{\rm dA}} \! = & \frac{e^2}{m^2\omega^2} \! \sum_{nl} M_{nn} \bigg[ \frac{p^+_{nl}p^-_{ln}(f_{l}-f_{n})}
{(E_l-E_n+i\hbar\Gamma_{ln}-\hbar\omega)^2} \\
 & ~~~~+\frac{p^-_{nl}p^+_{ln}(f_{l}-f_{n})}
{(E_n-E_l+i\hbar\Gamma_{nl}-\hbar\omega)^2}  \bigg], 
\nonumber
\end{split}
\end{equation}
\begin{equation}
\begin{split}
{\cal K}_{\rm dB} =&\frac{e^2}{m^2\omega^2} \! \sum_{nl}\frac{M_{nn}}{\hbar \omega} \frac{p^+_{nl}p^+_{ln}(f_{n}-f_{l})(i\hbar\Gamma_{ln}-\hbar\omega)}{ (E_l-E_n)^2+(\hbar\Gamma_{ln}+i\hbar\omega)^2} .
\label{eq:magnetization_expression}
\end{split}
\end{equation}
Here $p_{nl}$ are the matrix elements of the momentum operator, for relativistic electronic states $n\bm{k}$ with energies $E_{n\bm{k}}$, $\hat{p}^{\pm} = \hat{p}_x \pm i \hat{p}_y$, $f_n =f(E_{n\bm{k}})$ are the Fermi functions, $\hbar \omega$ is the laser photon energy, and $\Gamma_{nl}$ are linewidths. $M_{nm} = \mu_B (L_{nm}^z + 2S_{nm}^z)$ are magnetization matrix elements. Note that the the wavevector $\bm{k}$ is implicitly included.
The first term in Eq.\ (\ref{eq:magnetization_expression}) contains $M_{nm}$ off-diagonal in the electronic states ($n\neq m$); it corresponds to electronic Raman scattering. The other two terms, with diagonal matrix elements $M_{nn}$, correspond to Rayleigh scattering. For further details, see Ref.\ \cite{battiato14}.

Notably, the above formulation is applicable for both lossy and lossless media, and only requires the electronic states from \textit{ab initio} relativistic DFT calculations. Thus, our formalism takes into account the effect of light absorption, which, in all other previous models has been explicitly \cite{pitaevskii61,khorsand12}  or implicitly ignored \cite{pershan66,qaiumzadeh13,popova11}. However, since most of the recent measurements of optically induced magnetization were performed on strongly absorbing metals \cite{stanciu07,vahaplar12,hassdenteufel13,schubert14,mangin14,lambert14} it is imperative to take absorption into account \cite{anoteIFE}.


The laser-induced magnetization appears as a non-linear response to the laser field.  As the magnetization operator contains both orbital and spin contributions, we can use ${M}_{ind} = \mu_B \textrm{Tr} \{ \hat{L}^z\hat{\rho}^{[2]}\} +2 \mu_B \textrm{Tr}\{ \hat{S}^z \hat{\rho}^{[2]} \}$, to separate these, and  rewrite the total induced magnetization as  $M_{ind} = 2\textrm{Re}\,[{\cal K}(\omega ) E^2_0]
\equiv {K}^{\rm IFE} (\omega) I/c = [{K}^{\textrm{IFE}}_{L} (\omega) + {K}^{\textrm{IFE}}_S (\omega) ] I/c$, with $I$ the laser intensity $\varepsilon_0cE_0^2 /2$, $c$ the speed of light, and ${K}^\textrm{IFE} (\omega)=4\textrm{Re}\,[{\cal K} (\omega)]/\varepsilon_0$.
With this definition $K^{\textrm{IFE}}$ has the dimension of an inverse magnetic induction, T$^{-1}$.
The induced magnetic moment for a certain volume $V$ is easily obtained from $M_{ind}^{V}=M_{ind}V$.

In the following we employ 
electronic structure calculations to compute both ${K}^{\rm IFE}_{L}$ and  ${K}^{\rm IFE}_{S}$ for typical metals. 
Our relativistic calculations are valid for strong spin-orbit coupling (SOC).
We use the augmented spherical wave (ASW) method (see Ref.\ \onlinecite{eyert07}) which has already provided a very efficient relativistic bandstructure framework for accurate calculations of linear response functions, such as the conductivity tensor \cite{oppeneer01,oppeneer92}.
The evaluation of the momentum operator matrix elements ${{p}}_{nl}(\bm{k})$ and details of the $k$-space integration method can be found in Refs.\ \cite{oppeneer92,note,mondal15moke}. The linewidth $\Gamma$ is here assumed to be state independent. We use $\hbar \Gamma = 0.03$ Ry, a value which, in  \textit{ab initio} calculations of linear-order optical response tensors was found to give a good description for metals \cite{oppeneer01}.

\begin{figure}
\includegraphics[width=0.99\linewidth]{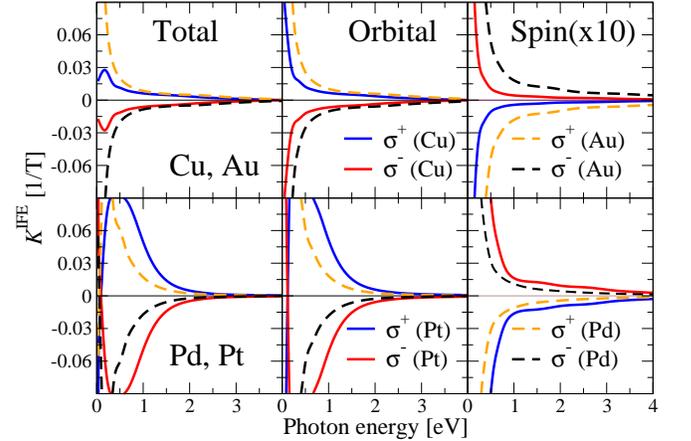}
\caption{(Color online) Calculated total, orbital, and spin IFE ${K}^{\textrm{IFE}}(\omega )$ as function of the photon energy for several nonmagnetic metals: Cu and Au (top) and Pd and Pt (bottom). The opposite circular laser polarizations are denoted with $\sigma^{\pm}$.}
\label{fig:f2}
\vspace*{-0.2cm}
\end{figure}


In Fig.\ \ref{fig:f2} we present the calculated spin, orbital, and total  $K^{\textrm{IFE}}$ functions for several nonmagnetic metals, Cu, Au, Pd, and Pt,  and for opposite laser polarizations, $\sigma^{\pm}$.  The calculations demonstrate that there exists both a large spin and an orbital IFE, 
which was not known previously,  and, furthermore,  that the small IFE of the spin component (shown $\times 10$) competes with that of the orbital component, as these two have opposite sign.
It can moreover be recognized that the IFE constants are materials specific:
For  Pd and Pt a larger total $K^{\textrm{IFE}}$ is obtained than for Cu and Au, particularly for photon energy $\hbar \omega \le 1$ eV.  As the expressions for the IFE components have a $1/\omega^2$ prefactor, see Eq.\ (\ref{eq:magnetization_expression}), the IFE functions typically diverge for $\omega \rightarrow 0$. For nonmagnetic metals all IFE components are fully \textit{antisymmetric} in the photon spin angular momentum,
i.e., the magnetization imparted with $\sigma^{-}$ helicity is precisely opposite to that induced with $\sigma^{+}$ helicity. 


\begin{figure}[tb]
\includegraphics[width=0.95\linewidth]{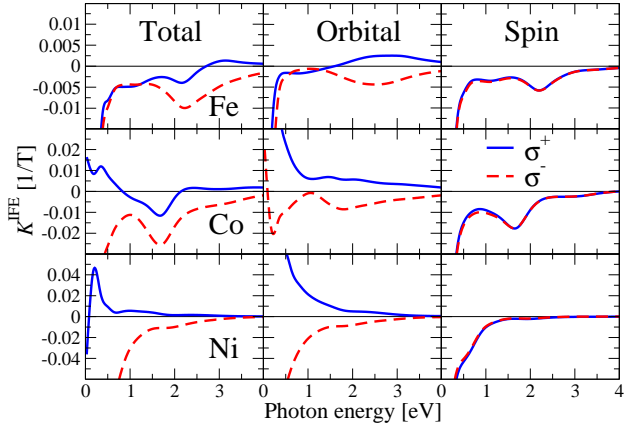}
\caption{(Color online) Calculated total, orbital, and spin inverse Faraday constant ${K}^{\textrm{IFE}}(\omega )$ as function of the photon energy and laser polarizations $\sigma^{\pm}$ for ferromagnetic bcc Fe, hcp Co, and fcc Ni.}
\label{fig:f3}
\end{figure}

Next, we have computed the laser-induced magnetization in the $3d$ ferromagnets, bcc Fe, hcp Co, and fcc Ni. The calculated $K^{\textrm{IFE}}$ functions are shown in Fig.\ \ref{fig:f3}. We find that the characteristics of laser-induced magnetization in ferromagnets are very different from those of nonmagnetic materials. 
The difference between spin and orbital components of the induced magnetization provides an even deeper insight into the physics of the IFE.
In contrast to the above nonmagnetic metals, the spin component $K^{\textrm{IFE}}_S$ is large, even dominant at optical frequencies, and moreover, it is not antisymmetric in the helicity. Surprisingly, $K^{\textrm{IFE}}_S$ does not even vary with the helicity, 
even though sizable values are obtained.
Notably, when we compute the IFE function of nonmagnetic bcc Fe (with zero exchange splitting, not shown), we find that its spin and orbital components are completely antisymmetric in the helicity.  The orbital contribution $K^{\textrm{IFE}}_L$ exhibits a different behavior from $K^{\textrm{IFE}}_S$ for the ferromagnets: $K^{\textrm{IFE}}_L$ is approximately antisymmetric in the helicity. This behavior  can be understood from the presence of the symmetry-breaking direction of the initial magnetization and hence full antisymmetry is not expected.  A consequence of the findings is that, as discussed further below, it will make a difference when one tries to induce optically a certain magnetization in a ferromagnetic metal or in a nonmagnetic material. Our results show that the previous common interpretation of the IFE, as an absorption-free, helicity antisymmetric effect, complementary to the circular magnetic dichroism present under absorbing conditions \cite{khorsand12}, has to be revised.

\begin{figure}[tb!]
\includegraphics[width=0.95\linewidth]{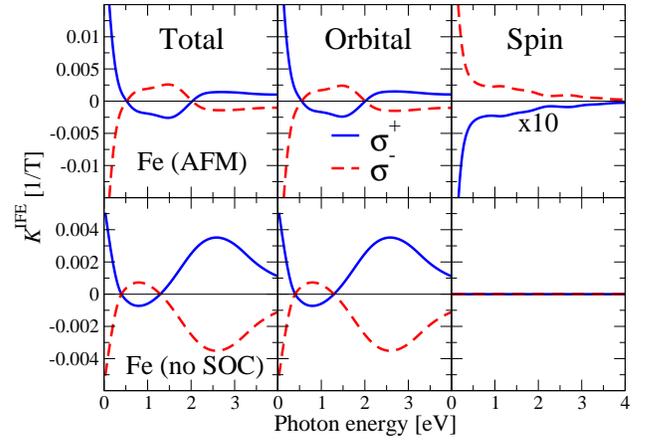}
\caption{(Color online) Calculated total, orbital, and spin IFE ${K}^{\textrm{IFE}}(\omega )$ as function of the photon energy for synthetic antiferromagnetic Fe (top panels) and for ferromagnetic Fe with zero spin-orbit coupling (bottom panels). A magnification of ten is used only for the top-right panel.
}
\label{fig:f4}
\end{figure}

All-optical helicity-related switching has been demonstrated in a variety of metallic materials, as e.g., rare-earth transition-metal ferrimagnets, synthetic antiferromagnets, and ferromagnets \cite{stanciu07,hassdenteufel13,schubert14,mangin14,lambert14}.
To investigate the influence of the magnetic structure on the IFE response we have computed the $K^{\textrm{IFE}}$ of synthetic antiferromagnetic Fe.
 The electronic structure of this artificial Fe 
 antiferromagnet has been simulated 
by aligning the moment of the Fe atom on the body centered position antiparallel to those on the corners of the bcc unit cell. The results for the $K^{\textrm{IFE}}$ function are shown in Fig.\ \ref{fig:f4} (top). In contrast to the ferromagnetic metals the induced IFE spin-component is antisymmetric and almost zero. The orbital IFE component does not vanish and is exactly antisymmetric in $\sigma^{\pm}$. We also note  that the energy dependence of the IFE of antiferromagnetic Fe is very different from that of ferromagnetic Fe. Our calculations thus show the induced magnetization to be antisymmetric in the helicity for nonmagnetic and antiferromagnetic materials, and we expect this property for paramagnetic materials as well.
 
Further insight in the origin of the IFE is gained by setting the spin-orbit coupling 
to zero in calculations for ferromagnetic Fe. The result of these calculations, given in Fig.\ \ref{fig:f4} (bottom)  evidence that the SOC is fully responsible for the imparted magnetization in the spin component.
The orbital component $K^{\textrm{IFE}}_L$ is however nonzero and antisymmetric in the helicity, even though the initial ferromagnetic spin polarization is nearly the same as that calculated with SOC included. 
The nonzero $K^{\textrm{IFE}}_L$ component can be traced to ${\cal K}_{\rm o}$, the first term of Eq.\ (\ref{eq:magnetization_expression}), which contains nonzero off-diagonal elements of the angular momentum operator, $L_{nm}^z$. This operator does not commute with the Hamiltonian, hence these elements do not vanish. As the effect of SOC is to mix the spin components of the bands, the SOC emerges as a key factor to obtain a large laser-induced magnetization.



\begin{figure}[tb!]
\includegraphics[width=0.95\linewidth]{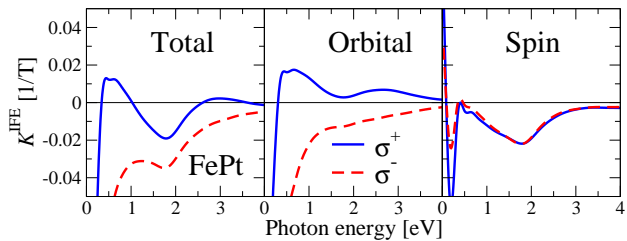}
\caption{(Color online) Calculated total, orbital, and spin IFE ${K}^{\textrm{IFE}}(\omega )$ of ferromagnetic iron-platinum as a function of the photon energy. }
\label{fig:f5}
\end{figure}

Before discussing the consequences of our \textit{ab initio} calculations we present calculated results for FePt. Iron-platinum is currently the main candidate for ultrahigh density magnetic recording \cite{weller13,stipe10}, because of its huge magnetocrystalline anisotropy energy, a property that enables stable magnetic bits of nanometer size. Lambert \textit{et al.}\ \cite{lambert14} recently demonstrated all-optical helicity dependent switching in FePt.
In Fig.\ \ref{fig:f5} we show the calculated results for FePt. We obtain an IFE dependence that is typical of a ferromagnetic material: $K^{\textrm{IFE}}$ is asymmetric in the helicity but not antisymmetric. We also observe that the $K^{\textrm{IFE}}$ curves show characteristics of both elemental Fe and Pt: the spin component of the IFE is very similar to the spin component on ferromagnetic Fe, while the orbital component is quite large and, with a near-antisymmetric behavior, similar to that of Pt.

In Table \ref{tab:mag_mag} we present calculated values of the induced total magnetization, assuming a laser frequency $\hbar \omega=1.55$ eV  and intensity of 10 GW/cm$^2$, values typically used in recent experiments \cite{vahaplar12,mangin14}. The induced magnetizations are of the order of $10^{-2} - 10^{-3}$ $\mu_B$ per atomic (Wigner-Seitz) volume, with the largest value obtained for FePt. Notably, for nonmagnetic metals the largest induced magnetizations are obtained for materials with large SOC, like Au and Pt. 
In previous work the IFE has not been treated as an induced magnetization, but as an effective optomagnetic field $B_{opt}$ \cite{vahaplar12,kimel05,kirilyuk13}. In this way the influence of the IFE on a magnetic material has been described as an effective Zeeman field $\bm{B}_{opt} \cdot \bm{M}_i$ acting on the  atomic spin moment $\bm{M}_i$ with unchanged length \cite{vahaplar12,qaiumzadeh13}. While this approach may provide insight in how all-optical helicity-dependent switching can occur, using e.g.\ spin-dynamics simulations \cite{vahaplar12,nieves16}, we note that the laser-induced spin and orbital magnetizations show a more complex behavior that cannot be captured with one field acting on the spin moments,  since its action on the orbital and spin components is quite different.
Nonetheless, to conform with the previous approach we have computed the effective Zeeman field that would be needed to induce the same magnetization as imparted by the IFE. The calculated values for $B_{opt}$ are also given in Table \ref{tab:mag_mag}. Although at first the values of the induced moments appear small, quite large fields of up to hundreds of Teslas are actually needed to generate a similar magnetization. The computed optomagnetic fields of Pd and Pt are smaller than those of Cu and Au, which is due to the fact that Pd and Pt with not completely filled $d$-shells are more susceptible to spin-polarization.

%
%
\begin{table}[ht!]
\begin{ruledtabular}
\vspace*{-0.3cm}
\caption{Calculated values of the optical helicity-dependent laser-induced magnetization (in $\mu_B$ per atomic volume), for various metals, assuming a typical photon energy of 1.55 eV and intensity of $10$ GW/cm$^2$. Also given is the computed optomagnetic field $B_{opt}$, i.e., the Zeeman field needed to induce the same magnetization as the circularly polarized laser field.}
\label{tab:mag_mag}
\begin{tabular}{l c c c c}
Material & \multicolumn{2}{c}{$M_{\rm ind}$ ($\times 10^{-3}\mu_B$/at. vol.)} & \multicolumn{2}{c}{$B_{\rm opt}$ (Tesla)} \\
\line(1,0){60} & \multicolumn{2}{c}{\line(1,0){115}} & \multicolumn{2}{c}{\line(1,0){60}}\\
\vspace{0.5mm}
Ferromagnetic              &~~~~~~$\sigma^+$  &  ~~~~~~$\sigma^-$ &  ~~~$\sigma^+$  &  $\sigma^-$ \tabularnewline
\hline
Ni (fcc) & ~~~~$+1.4$   & ~~~~$-4.5 $ & ~$+16$ & $-50$\\
Fe (bcc) & ~~~~$-3.3$  & ~~~~$-5.5 $ & ~$-40$ & $-65$\\
Co (hcp) & ~~~~$-4.8$ & ~~~~$-13$ &  ~$-100$ & $ -260$\\
FePt     & ~~~~$-15$  & ~~~~$-33$ &  ~$-190$ & $-420$\\
\vspace{-0.1cm}
\line(1,0){60} & \multicolumn{2}{c}{\line(1,0){115}} & \multicolumn{2}{c}{\line(1,0){60}}\\
\vspace{0.5mm}
Nonmagnetic & \multicolumn{2}{c}{~$\sigma^+/\sigma^-$} &  \multicolumn{2}{c}{~$\sigma^+/\sigma^-$} \\[0.1cm]
\hline
Cu (fcc) & \multicolumn{2}{c}{$\pm2.0$}& \multicolumn{2}{c}{$\pm100$}\\
Pd (fcc) &  \multicolumn{2}{c}{$\pm2.7$}&  \multicolumn{2}{c}{$\pm2$}\\
Pt (fcc) &  \multicolumn{2}{c}{$\pm6.5$}&  \multicolumn{2}{c}{$\pm28$}\\
Au (fcc) &  \multicolumn{2}{c}{$\pm7.5$}&  \multicolumn{2}{c}{$\pm350$}\\
\end{tabular}
\end{ruledtabular}
\vspace*{-0.3cm}
\end{table}

Our calculations explain puzzling aspects of all-optical magnetization switching experiments \cite{vahaplar12,lambert14,hadri16,hadri16a}.  All-optical magnetization reversal triggered by \textit{single} laser pulses was first discovered for GdFeCo alloy near its compensation point \cite{vahaplar12,stanciu07,schubert14}. Interestingly,  the effect was found to be almost helicity-independent, i.e., opposite helicities act in a very similar way. This can be understood from our results for absorbing ferromagnetic materials, where opposite helicities can induce quite similar magnetizations. Near the compensation point both helicities will hence exert a similar effect on the nearly compensating magnetic moments and thus initiate helicity-independent switching.
Intriguingly, this behavior of ferrimagnets is different from that found in earlier IFE experiments on paramagnetic materials \cite{vanderziel65} which suggested the IFE to be antisymmetric in the helicity, but it is consistent with our \textit{ab initio} calculations that predict an antisymmetric IFE for nonmagnetic and antiferromagnetic materials but not for ferro- or ferri-magnetic materials.
Recently, all-optical helicity-dependent switching was observed for a variety of metallic materials including synthetic ferrimagnets and ferromagnets \cite{hassdenteufel13,mangin14,lambert14,hadri16}. 
To achieve switching in these experiments \textit{repeated} laser pulses were applied \cite{lambert14,hadri16,hadri16a}. This ``two-step" mechanism  appearing after multi-shot laser pulses \cite{hadri16a} can be understood from our calculations as well: the initial laser pulses lead to demagnetization and spin moment disorder, in a first step.
For ferro- or ferri-magnets this initial process will again be alike for both laser helicities. Once the magnetic order in the material is largely quenched (and, being similar to paramagnetic) the opposite helicities of the subsequent laser pulses will induce IFE magnetizations with opposite sign, which, in a second step, will push the re-magnetization process in one or the other direction, consistent with recent experiments \cite{hadri16a}. 

To engineer optimal conditions for helicity-dependent magnetization reversal, our \textit{ab initio} calculations suggest the following guidelines:
first, tuning the laser frequency, in particular to lower frequencies may enhance the induced magnetization. Second, materials with a strong spin-orbit coupling will have an enhanced IFE caused by the stronger mixing between the spin-up and spin-down spinor states. Third, magnetization reversal can be more easily reached when the action of the circular polarized laser is antisymmetric in the helicity, as e.g.\ for antiferromagnets. However, as  antiferromagnetic materials are not suitable for recording applications, ferrimagnetic or even ferromagnetic materials can be used but in combination with repeated laser pulses.

To summarize, our \textit{ab initio} theory unveils unexpected features of the magnetizations imparted by circularly polarized laser light in absorbing metals. The common interpretation that the underlying IFE is an absorption-free, helicity antisymmetric effect has to be revised. Both spin and orbital magnetizations are induced by the polarized laser light, yet these display a very different dependence on frequency, helicity, and on the material's magnetic order. Notably, our results stress that \textit{ab initio} calculations provide a suitable materials specific platform for accomplishing all-optical magnetization control.


We thank  U.\ Nowak, P.\ Maldonado, M.\ M{\"u}nzenberg, and S.\ Mathias for valuable discussions. 
This work has been supported by the European Community's Seventh Framework Programme 
under grant agreement No.\ 281043, FemtoSpin, the Swedish Research Council (VR), the R{\"o}ntgen-{\AA}ngstr{\"o}m Cluster, the K.\ and A.\ Wallenberg Foundation and the Swedish National Infrastructure for Computing (SNIC).
\bibliographystyle{apsrev4-1}

\begin{thebibliography}{31}
\bibitem [{ano()}]{anoteIFE}%
  \BibitemOpen
  \href@noop {} {}\bibinfo {note} {In the often used model of Pitaevskii
  \cite{pitaevskii61} the IFE is assumed to be absorption free and fully
  antisymmetric in the laser helicity.}\BibitemShut {Stop}%
\bibitem [{not()}]{note}%
  \BibitemOpen
  \href@noop {} {}\bibinfo {note} {For the Brillouin zone integration we used
  414720 $k$-points in the double irreducible wedge for Ni, Fe, Cu, Pd, Pt, and
  Au, and 491520 $k$-points in the quadruple irreducible wedge for hcp Co, FePt
  and AFM Fe.}\BibitemShut {Stop}%
\end{thebibliography}

\end{document}